\documentclass[aps,prl,twocolumn,showpacs,superscriptaddress,floatfix,nofootinbib]{revtex4}
\usepackage{graphicx} 
\usepackage[english]{babel}
\usepackage{amssymb}
\usepackage{amsmath}
\begin{document}
\newcommand{\p}{\prime}
\newcommand{{\ud}}{{\phantom1}}  
\newcommand{\bh}{\mbox{\scalebox{.9}{$\bullet$}}}  
\newcommand{\bb}{\mbox{\scalebox{0.5}{$\blacksquare$}}}  
\newcommand{\bd}{\mbox{\rotatebox{45}{\scalebox{0.5}{$\blacksquare$}}}}  
\title{On the origin of duality in the quantum Hall system}
\author{C.A. L\"utken}
\affiliation{Theory Group, Department of Physics, University of Oslo}
\affiliation{Theory Division, CERN, CH-1211 Geneva 23}
\author{ G.G. Ross}
\affiliation{Theory Division, CERN, CH-1211 Geneva 23}
\affiliation{Rudolf Peierls Centre for Theoretical Physics, Department of Physics,
University of Oxford}
\date{\today}
\preprint{CERN-PH-TH-2010-174}
\preprint{OUTP-10-18P}
\begin{abstract}
We discuss the possible origin of the  duality observed in the quantum Hall current-voltage characteristics. 
We clarify the difference between ``particle-vortex"  (complex modular) duality, which acts on the 
full transport tensor,  and  ``charge-flux" (``real") duality, which acts directly on the filling factor. 
Comparison with experiment strongly favors the form of duality which descends from the modular 
symmetry group acting holomorphically on the compexified conductivity.
\end{abstract}
\maketitle

\begin{figure}[t]
\begin{center}
\includegraphics[scale = .8]{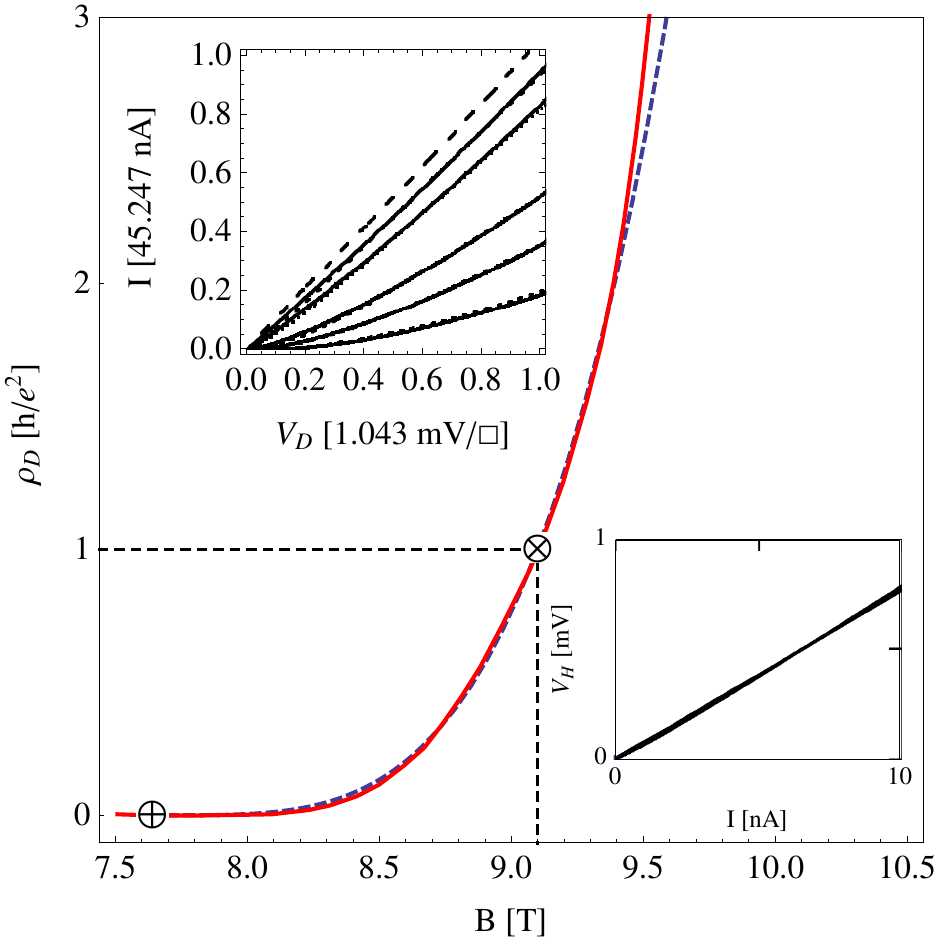}
\end{center}
\caption[fittedrho]{(Color online) Experimental discovery of duality in the QH system (adapted from ref.\,\cite{Tsui1}).
The solid red curve is a reproduction of the lowest temperature resistivity curve $\rho_D^\ud(B)$,obtained at $T = 26 \,mK$.
The dashed black curve shows $\rho_D^{\rm fit} (B) = (\Delta B/\Delta B_\otimes)^a$, fitted to the data with $a = 3.8$.
\emph{Top inset:} Five pairs of $V_D^\ud(I)$-traces recorded on opposite sides of the liquid-insulator quantum phase transition, 
obtained at  $T = 21\,mK$ and $B$-field values in  the range $8.5 - 10.1\,T$.
The low field data $(\nu = 1/3)$ have been reflected in the symmetry axis $V_D = I$
and are practically indistinguishable from the high field data, showing that the five dual pairs of IV-characteristics are 
mirror images (inverse functions).
\emph{Bottom inset:}  $V_H^\ud(I)$-traces recorded in this experiment, 
with $B$-field values incremented in steps of $0.1\,T$ from $8.7\,T$ to $9.7\,T$.}
\label{fig:Probes1-Fig1}
\end{figure}

In a remarkable experiment\footnote{The Hall bar is a high mobility $(\mu = 5.5 \times 10^5\,cm^2/sV)$,
low density $(n = 6.5 \times 10^{10} \,/cm^2)$ GaAs/AlGaAs hetero-structure of size
$L_x \times L_y$, aligned with a current $I$ in the $x$-direction so that $R_{*x} = (L_*/L_y) \rho_{*x}$.  
The Hall resistance $R_H = \rho_H = \rho_{yx}$ is quantized in the fundamental unit  of resistance, 
$h/e^2 \approx 25.8 \;k\Omega$, while the dissipative resistance $R_D = \rho_{xx}/\square$  
is rescaled by the aspect ratio $\square = L_y/L_x$.
In this experiment $R_D^c \approx 23 \;k\Omega$ and  $\rho_{xx}^c  \approx 1\;\;[h/e^2]$\,\cite{Tsui2}.}    
Shahar et al.\,\cite{Tsui1} found clear evidence for a duality between current-voltage (IV) characteristics 
obtained on opposite sides of the quantum Hall (QH) liquid to insulator transition.  
They identified six pairs $(B, B_d)$ of ``dual" magnetic field values,
with $B$ and $B_d$ on opposite sides of the quantum phase transition at\footnote{Since no errors are given
in ref.\,\cite{Tsui1}, our least biased estimate is to take the largest error consistent with the published value: 
$B_c = 9.1\pm 0.05\,[T]$, and similarly for the other values of $B$.} $B = B_c\approx 9.1\,T$, 
separating the $\nu = 1/3$ fractional QH liquid from the QH insulator phase ($\nu = 0$).
With $B$-values in this range the transverse IV-characteristics $V_y(B,I) = R_{yx}(B,I) \cdot I$ 
was found to be independent of $B$ and linear in $I$, with slope  $R_H = R_{yx} \approx 3\;[h/e^2]$
(Fig.\,\ref{fig:Probes1-Fig1}, bottom inset). 
The critical value of the Hall resistivity is therefore $\rho_H^c = \rho_{yx}(B_c) \approx 3\;[h/e^2]$.  
The dissipative IV-characteristics $V_x(B,I) = R_{xx}(B,I) \cdot I$, on the other hand, are extremely non-linear in both phases, 
degenerating to a linear (Ohmic) relation only when $B\rightarrow B_c$  
(Fig.\,\ref{fig:Probes1-Fig1}, top inset).

Shahar et al.\,\cite{Tsui1} discovered that to each $B$ there exists a dual field value $B_d$, such that the dissipative IV-curve 
$V(B,I)$ (suppressing the now superfluous subscript on $V_x$) after reflection in the diagonal $V = I$ is virtually identical to 
the dual IV-curve $ V_d(B_d,I_d)$ in the opposite phase.   This is implies that
\begin{equation}
\rho_D^d(B_{d},I_d) = 1/\rho_D^\ud(B,I)\;\; ,
\label{eq:expRRd}
\end{equation}
a remarkable relation providing unambiguous evidence of a duality symmetry in the QH system, for a particularly simple 
quantum phase transition. It reveals a profound connection between the transport mechanisms deep inside the 
quantum liquid and insulator phases, a symmetry which holds to great accuracy even far from the quantum 
critical point found at:
\begin{equation}
(\rho_H^c, \rho_D^c)  \approx (3,1)\;\;[h/e^2]\; ,
\label{eq:expQCP}
\end{equation}
consistent with the self-dual value of eq.\,(\ref{eq:expRRd}). 

We are only aware of two (related) theoretical frameworks that aspire to account for these data (and which motivated 
the experimental investigation of duality).  The first\,\cite{Oxford1} is based on an effective description of the 
macroscopic theory that posesses a holomorphic modular symmetry relating the complexified response functions
\begin{eqnarray*}
\sigma = \sigma_{xy} + i\sigma_{xx} &=& {\phantom -}\sigma_H^\ud + i \sigma_D^\ud \\
\rho = \rho_{xy} + i\rho_{xx} &=& - \rho_H^\ud + i \rho_D^\ud
\end{eqnarray*}
in different QH phases. 
It successfully predicts the full phase diagram of the QH system, both integer and fractional, including the position of the quantum critical points governing the scaling behaviour of transitions between QH levels\,\cite{Oxford1}. 
A ``microscopic"\footnote{Our timid use of the term ``microscopic" signals that a rigorous derivation of these models from the quantum electrodynamics of disordered media is far beyond current theory.  
For each model assumptions about the effective degrees of freedom at some intermediate scale 
are made.} interpretation of the symmetry as ``particle-vortex duality" was given in ref.\,\cite{Cliff}. 

The second framework is provided by a ``microscopic" model\,\cite{KLZ} in which a so-called ``flux attachment transformation" 
maps the two-dimensional electron system in a magnetic field onto a bosonic system in a different ``effective" field. 
This ``charge-flux duality" resulted in a set of rules, known collectively as ``the law of corresponding states", 
which relates QH states carrying different filling factors $\nu$. It also determines the topology of  the phase diagram, 
but neither the location of quantum critical points nor the geometry of renormalization group (RG) flows in the complex 
conductivity plane are obtained by this method.

In refs.\,\cite{Tsui1,nu_duality}  the law of corresponding states was used to determine the pairs of filling factors $(\nu,\nu_{d})$ 
that correspond to states related by charge-flux (particle-vortex) duality. 
For the $\nu = 0$ to $\nu = 1$ transition the corresponding states are related by particle-hole duality giving $\nu_d = 1 - \nu$. 
The flux attachment transformation $1/\nu^{\prime} = 1/\nu + 2 m$ ($m$ integer) maps this transition to the $\nu = 0$ to $\nu = 1/k$ 
transition, with $k = 2 m + 1$, giving the duality relation:  $1/\nu_d - k = (1/\nu - k)^{-1}$. 
For the $\nu = 0$ to $\nu = 1/3$ transition this gives
\begin{equation}
\nu_{d}(\nu)  = \frac{1 - 3\nu}{3 - c_\nu \nu}\quad (c_\nu = 8)\;,
\label{eq:vvd}
\end{equation}
which relates the filling $\nu = \nu(B)$ to a dual filling $\nu_d = \nu(B_d)$.

The self-dual point of this transformation is $\nu_* = 0.25$, distinct from the critical value  $\nu_c \approx 0.28$ found 
experimentally\,\cite{Tsui2}.  In fact, this argument fails to reproduce any of the main features observed experimentally.
It has nothing to say about  the Hall response of the system, including the fact that it remains constant at its critical value 
$\rho_H^c \approx 3\,[h/e^2]$ across the transition. 
It is also not clear that eq.\,(\ref{eq:expRRd}) relating dual values of $\rho_D^\ud$ applies for the the dual pairs of 
filling factors given by eq.\,(\ref{eq:vvd}), because the mapping used to derive the latter strictly applies only on 
the plateaux where $\rho_D^\ud$ vanishes. Also, as discussed below, it does not reproduce the measured values of the dual filling factors. 

The situation is quite different for the modular symmetry\,\cite{Oxford1}.  
The transformations relating corresponding states
are fractional linear (M\"obius) maps that move points on the upper half of the complex 
$\rho$ (or $\sigma$) plane ($\rho_D^\ud$ and $\sigma_D^\ud$ are positive). 
It swaps the entire quantum liquid phase,  a region of the $\rho$ (or $\sigma$) plane ``attached"
only to $\rho_\oplus^\ud = - \rho_H^\ud = - 3$ ($\sigma_\oplus = \sigma_H^\ud = 1/3$) on the real axis,\footnote{In other words, 
the domain of attraction (universality class) of the RG flow towards the real infra-red fixed point $\rho_\oplus$.}
with the entire QH insulator phase,  a region of the  $\rho$ (or $\sigma$) plane attached only to  
$\rho_\oplus^\ud  = i\infty$  ($\sigma_\oplus  = 0$).
Since these transformations act on complex coordinates rather than on the real filling factor $\nu$, 
the prediction for the dual pairs is different.

The particular modular transformation  that maps the (infrared) plateaux fixed points 
$\rho_\oplus^\ud = -3$ and $\rho_\oplus^\ud = i\infty$ into each other has
 the form:
\begin{equation}
\rho^{d}(\rho)  = - \frac{c + 3\rho}{3 + \rho}\;.
\label{eq:rrd}
\end{equation}
This transformation is modular iff $\det \rho^{d} = 1$, which fixes $c = 10$.
The self-dual point, located at  
\begin{equation}
(\rho_H^*, \rho_D^*) = (3,1)\;\;[h/e^2]\;,
\label{eq:modQCP}
\end{equation}
is an RG fixed point and therefore the quantum critical point of this phase transition\,\cite{Oxford1}.
It coincides exactly with the experimental finding recorded in eq.\,(\ref{eq:expQCP}).

Moreover, starting from the plateau corresponding to $\nu = 1/3$, at sufficiently low temperature the dynamics 
associated with the modular symmetry is expected to force the flow along the line $\rho_H^\ud = 3$\,\,\cite{Cliff2}. 
This constraint is certainly satisfied experimentally with great accuracy, as seen in the bottom inset in Fig.\,\ref{fig:Probes1-Fig1},
copied from ref.\,\cite{Tsui1}.   All the $V_H^\ud(I)$-traces recorded this experiment, at eleven different values of the magnetic field, are indistinguishable, with slope $dV_H/dI \approx \rho_H^\ud \approx 3 [h/e^2]$.

With this constraint the modular duality transformation of the longitudinal component 
following from  eq.\,(\ref{eq:rrd}) is:
\begin{equation}
\rho_D^d  = 1/\rho_D^\ud\;\; .
\label{eq:modRRd}
\end{equation}
Thus it is a \emph{prediction} of modular symmetry that there should exist dual pairs of $B$-fields for which
the experimental result recorded in eq.\,(\ref{eq:expRRd}) should hold with great precision, i.e., at the same level of accuracy 
that the emergent symmetry holds.  
In a transition to the insulator phase there must of course be pairs of points with inverse values of the resistivities,
but duality implies that the physics of the system at dual parameter values is the same, 
explaining why the full nonlinear structure of the IV-curves evident in Fig.\,{\ref{fig:Probes1-Fig1}  coincide so precisely. 

To gain further insight into the origin of the differences between the two approaches it is instructive to consider the 
structure in the conductivity plane. The modular symmetry transformation swapping the plateaux fixed points 
 $\sigma_\oplus = 0$ and $\sigma_\oplus = 1/3$ follows immediately from  eq.\,(\ref{eq:rrd}):
\begin{equation}
\sigma^d(\sigma)  = \frac{1 - 3\sigma}{3 - c \sigma}\;\;. 
\label{eq:ssd}
\end{equation}
This is superficially similar to eq.\,(\ref{eq:vvd}), but these two transformations are in fact very different.
Not only does modularity ($\det \sigma^d = 1$) fix $c = 10$, but more importantly eq.\,(\ref{eq:ssd}) is a 
much stronger statement about emergent global symmetries in the QH system than eq.\,(\ref{eq:vvd}).

Decompressing the complex transformation in eq.\,(\ref{eq:ssd}) into components gives two non-linear functions
$\sigma^d_H (\sigma_H^\ud, \sigma_D^\ud)$ and $\sigma^d_D (\sigma_H^\ud, \sigma_D^\ud)$ that
in general are inextricably entangled.  
The constraint $\rho_H^\ud = 3$  restricts the flow  to the semi-circle $\vert\sigma\vert^2 =\sigma_H^\ud/3$, 
giving the constrained modular duality transformation of interest here:
\begin{equation}
\sigma^d_H  = \frac{1 - 3 \sigma_H^\ud}{3 - c_m \sigma_H^\ud}\quad (c_m = 80/9)\;.
\label{eq:ssdxy}
\end{equation}
Since $\sigma^d_H$ now depends only on $\sigma_H^\ud$, this is as close to a formal similarity to the 
$\nu$-duality in eq.\,(\ref{eq:vvd}), following from the law of corresponding states, as we can get.
However, since $\sigma_H^\ud$ is only given by the filling factor on the plateaux, and $c_m$ is quite different from $c_\nu$,
there is no way to reconcile these transformations.

In order to exhibit a more quantitative comparison with the data we show in Fig.\,\ref{fig:Probes1-Fig2}\,(a) the observed 
values for  $\Delta\nu = \nu - \nu_c$ reported in ref.\,\cite{Tsui1}.  The boxed values are the images of the low-field data
(the black bullets, obtained in the $\nu = 1/3$ phase) under the transformation following immediately from eq.\,(\ref{eq:vvd}):
\begin{equation}
\Delta\nu_{d}(\Delta\nu)  = - \frac{(1 - 6\nu_c + 8\nu_c^2) + (8\nu_c - 3)\Delta\nu}{(8\nu_c - 3) + 8\Delta\nu}\;.
\label{eq:dvdvd}
\end{equation}
Clearly the image points $\Delta\nu_{\mbox{\scalebox{0.5}{$\Box$}}} =  \Delta\nu_{d} (\Delta\nu_{\bullet})$ 
do not coincide with the insulator fillings $\Delta\nu_{\circ}$ reported in ref.\,\cite{Tsui1}.

\begin{figure}[t]
\begin{center}
\includegraphics[scale = .8]{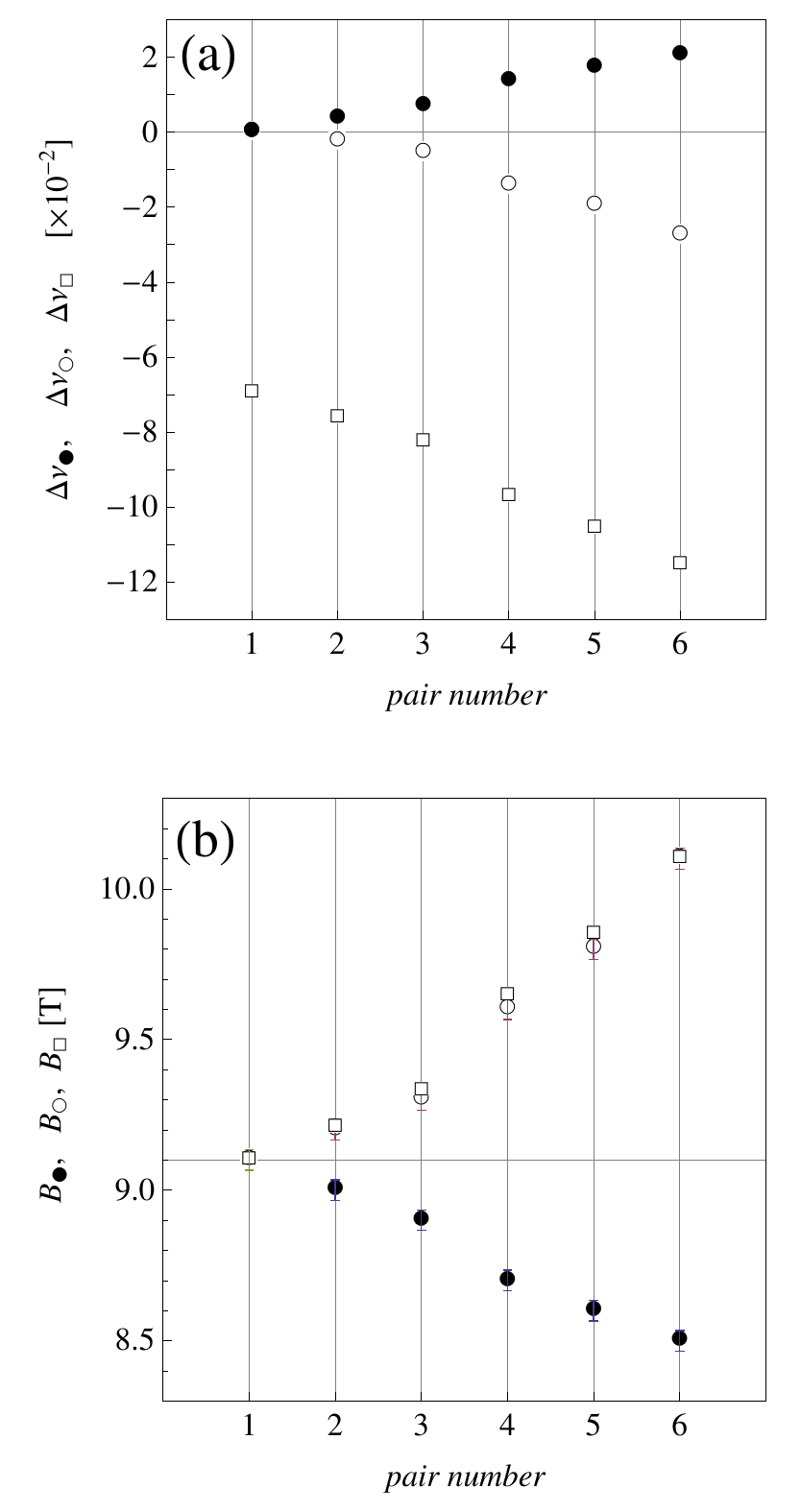}
\end{center}
\caption[endoflaw]{Black and white bullets  are the dual pairs
(a) $(\Delta\nu_{\bullet}, \Delta\nu_{\circ} )$ and (b) $(B_{\bullet}, B_{\circ} )$ (with error-bars discussed in the text),
reported in ref.\,\cite{Tsui1}.  According to the law of corresponding states the open icons in (a) should coincide.
According to the approximate duality relation in  eq.\,(\ref{eq:BBd}) the open icons in (b) should coincide.}
\label{fig:Probes1-Fig2}
\end{figure} 

By comparison, the full non-linear content of modular duality is contained in eq.\,(\ref{eq:rrd}), 
which is in perfect agreement with the experiment (within experimental accuracy), as already discussed.
The prediction following from the emergent symmetry is that there should exist dual parameter values so that the 
full non-linear resistivites collapse as shown in Fig.\,\ref{fig:Probes1-Fig1}.    
In short, the experimentally observed duality is in clear disagreement with the form obtained using 
the law of corresponding states, and in excellent agreement with the structure predicted by the modular symmetry $\Gamma_{\rm H}$.

The modular symmetry by itself does not predict the values of the dual magnetic fields. 
However, we expect that the relation between $B_{d}$ and $B$ should, to a good approximation, 
be fractional linear (a Pad{\'e} approximant of order (1,1)) of the form 
 \begin{equation}
 B_{d}(B)=\frac{aB-b}{B-c}\;\;,
 \end{equation}
 with $a,b$ and $c$ are constants. This form has a simple pole at $B=c$ corresponding to the need to have 
 dual field values deep in the insulator phase. It also has the property that its form is maintained when expressing 
 $B$ as a function of $B_{d}$ as is required from the duality property of the theory. We can fix the three constants 
 by requiring that the self dual point should be at $B=B_{\otimes}$, and by demanding that the 
duality relation swap the plateau value, $B=B_{\oplus}\;(\nu=1/m)$, with the insulator value,  $B=\infty\;(\nu = 0)$.
The resulting duality relation  has the form 
\begin{equation}
B_d(B) =\frac{B B_\oplus  + (B_\otimes - 2B_\oplus) B_\otimes}{B - B_\oplus}\;\;.
\label{eq:BBd}
\end{equation}

 Both the parameters $B_\otimes$ and $B_\oplus$ are of course non-universal, 
but can be extracted from each experiment as follows.
$B_\otimes \approx B_\times$ is the critical field value, identified in the experiment from the 
temperature independent crossing point of all the resisitivity traces.  
The value of $B_\oplus$ can be directly obtained from the measured $B$ value at the centre of the plateau.

For the case of the transition from the $\nu=1/3$ plateau $m=3$. In the experiment discussed above\,\cite{Tsui1}  
 $B_\otimes = 9.1\,T$, and we estimate that $B_\oplus \approx 7.6\,T$, corresponding to the centre of the $\nu=1/3$ plateau. 
(For $\nu\propto B^{-1}$ this value agrees very well with the positions of the other plateau values.)

Fig.\,\ref{fig:Probes1-Fig2}\,(b) shows the dual pairs of field values found 
experimentally\,\cite{Tsui1} as black and white bullets.  
The boxed values are the images of the low-field data (the black bullets, obtained in the $\nu = 1/3$  phase) 
under the transformation in eq.\,(\ref{eq:BBd}).
The image points $B_{\mbox{\scalebox{0.5}{$\Box$}}} =  B_d( B_{\bullet})$ 
fall on top of the insulator values  $B_{\circ}$ reported in ref.\,\cite{Tsui1}, 
even in this linearized approximation.

One way to see why this linear approximation gives such a good result
is to observe that a simple power law gives an accurate model of the data.  The function:
\begin{equation}
\rho_D^{\rm fit} (B) = (\Delta B/\Delta B_\otimes)^a \;\;,
\end{equation}
with $\Delta B = B - B_\oplus$, $\Delta B_\otimes = B_\otimes - B_\oplus$ and the fitted value $a = 3.8$,
is shown as the dashed curve in Fig.\,\ref{fig:Probes1-Fig1}. We see that it gives a good fit to the real resistivity trace
(solid red curve, adapted from the lowest temperature trace in ref.\,\cite{Tsui1}) for a reasonable range of $B$-fields,
and only deviates when $B$ goes deeply into the insulator phase $(B \rightarrow \infty)$.
If we set $\rho_D^{\rm fit} (B_d) = 1/\rho_D^{\rm fit} (B)$, the fitting parameter drops out and eq.\,(\ref{eq:BBd}) follows.

Since duality relates all the plateau-insulator transitions, we expect the fractional linear approximation in  eq.\,(\ref{eq:BBd})
to work equally well for any transition of this type. It should therefore provide an accurate prediction for dual pairs of 
$B$-fields across any insulator-plateau  $(\nu = 0\leftrightarrow\nu = 1/m)$  quantum phase transition.

In summary, there is a significant difference between the duality predictions derived from the law of corresponding states, 
and the predictions derived from the modular symmetry acting holomorphically on the complexified conductivity or resistivity. 
While the former is strongly disfavoured by experiment, the latter gives a detailed explanation for the striking features of duality 
observed in the experiment, including the exact agreement (within experimental accuracy) of  the modular predictions 
in eqs.\,(\ref{eq:modQCP}) and (\ref{eq:modRRd}), with the data recorded in eqs.\,(\ref{eq:expQCP}) and (\ref{eq:expRRd}).
This provides further\,\cite{Oxford1} strong evidence for an emergent modular symmetry in the QH system. 

Since the two proposed emergent symmetries are distinct, effective field theories possessing these symmetries must also be different.
The nature of the effective degrees of freedom, i.e., the composite quasiparticles (possibly anyons), 
at large in the QH system is still an open problem, three decades after the discovery of the QH effect.  
We have explained how duality experiments are well suited to inform this issue, by efficiently discriminating between
candidate models dependent on specific degrees of freedom.

To date we are not aware of any experimental study of another plateau-insulator transition that probes  the duality relations in the quantum regime where the temperatures are low enough for the fixed point structure to be reached\,\cite{Oxford2}. Given that the duality predictions following from modular symmetry are rigid and precise, such measurements are capable of providing a definitive test of the emergent modular symmetry in the quantum Hall system.

\begin{figure}[t]
\begin{center}
\includegraphics[scale = .4]{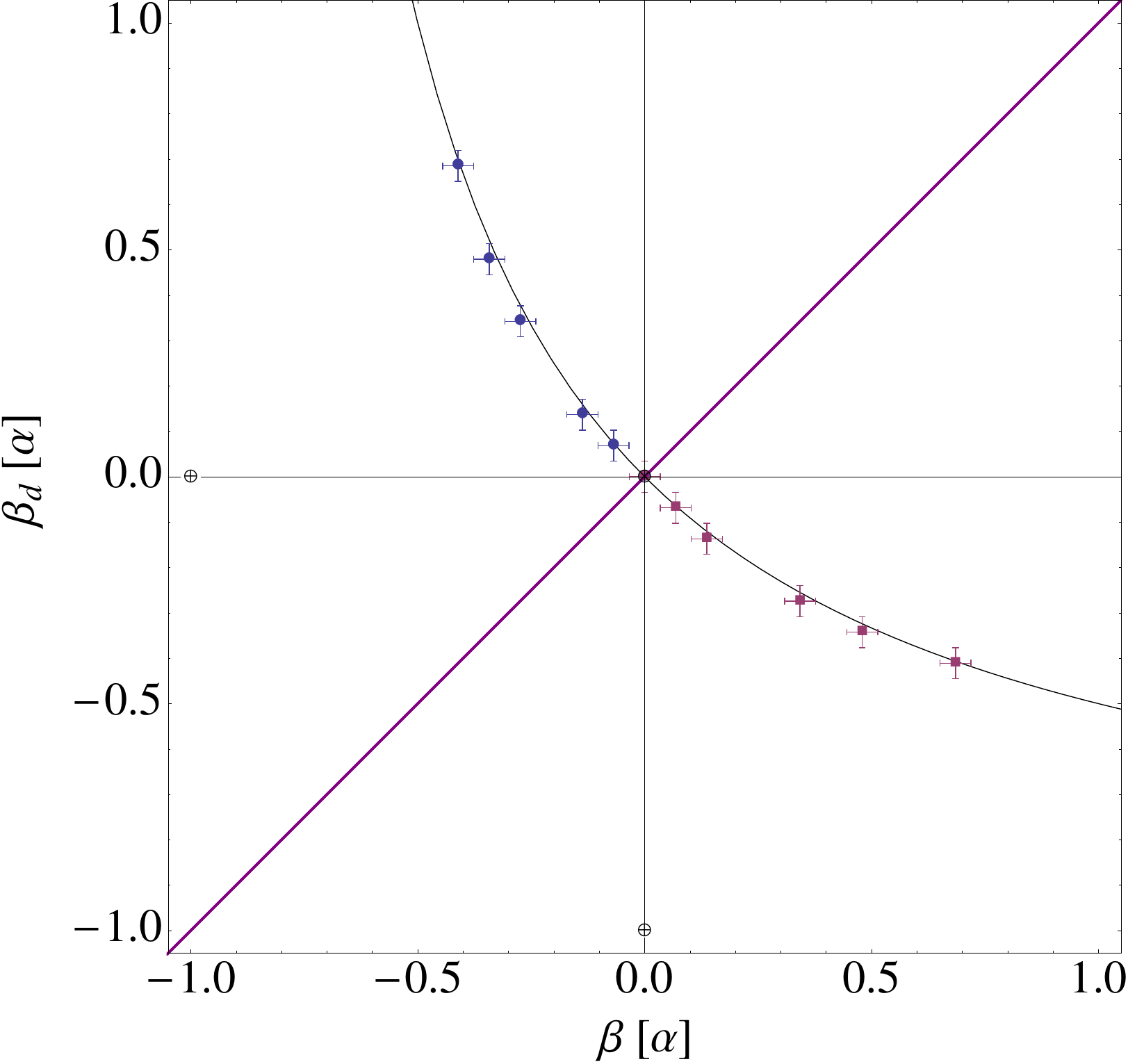}
\end{center}
\caption[endoflaw]{(Color online) Universal curve for $B$-field duality across any plateau-insulator transition, 
at leading order in the Pad\'e-expansion.  The only available duality data on $B$-fields is taken from ref.\,\cite{Tsui1}.}
\label{fig:Probes1-Fig3}
\end{figure} 

\emph{Epilog:}  We can write the $B$-duality relation in a universal form that can be easily tested when duality is 
investigated for  other plateau-insulator  transitions.
Start by introducing the ``reduced" $B$-field $b = (B - B_\otimes)/B_\otimes$, which is analogous to the reduced temperature 
$t = (T -T _c)/T_c$ used in the study of classical phase transitions.
We also choose to measure all fields in the unit $\alpha =  - b_\oplus = (B_\otimes - B_\oplus)/B_\otimes > 0$,
whence $b = \beta\alpha \simeq \beta$ and  $b_d = \beta_d\alpha \simeq  \beta_d$.
With this notation all non-universal (system- and transition-specific) data have been absorbed in the units, 
and the duality transformation takes the universal form: 
\begin{equation}
\beta_d(\beta) = - \frac{\beta}{1 + \beta}\;\;.
\end{equation}
The prediction is now that the dual $B$-field data for any plateau-insulator transition will collapse onto this curve, shown
in Fig.\,\ref{fig:Probes1-Fig3}, where we have also added the only available duality data on $B$-fields\,\cite{Tsui1}.


\end{document}